# Long Term Logarithmic Annealing in p-MNOS RADFETs and Renormalization of Relaxation Parameters

P. A. Zimin, E. V. Mrozovskaya, V. A. Yushkova, V. S. Anashin, P. A. Chubunov, G. I. Zebrev

*Abstract* — **It was experimentally shown that annealing (fading) of the pMNOS based dosimeters has close to logarithmic temporal dependence during and after irradiation. The results are shown to be consistent with the previously proposed model.**

## I. INTRODUCTION

The RADiation senitive Field Effect Transistor (RADFETs) are the total-ionizing-dose sensors in which the threshold voltage shift is a quantitative indicator of the absorbed dose. [1].The p-channel stacked p-MNOS (Metal - Nitride – Oxide) structures are often used in these devices. The RADFETs must meet two fundamental dosimetric demands: a good compromise between dose sensitivity and stability with time after irradiation. The dose-rate effects impact affect both of these aspects. The dose rate effects can be divided into the true dose rate effects and the simultaneous annealing effects [2]. The first type of effects is caused by the dependence of the effective charge yield on dose rate. We investigated and discussed these effects, typical for thick insulators, in Ref. [3, 4]. This work aims to investigate the annealing effects in RADFETs with relatively thin gate insulators.

## II. EXPERIMENTS AND SIMULATIONS

### A. Experimental conditions

We investigated the stacked MNOS-based devices (see Fig. 1). The thicknesses of the $Si_3N_4$ and $SiO_2$ layers were of 45 nm. The MNOS devices were irradiated with various dose rates from 1 rad(Si)/s to 100 rad(Si)/s using a Co-60 source. The experimental dose error did not exceed 20%. Irradiation was performed at various temperatures (-40°C, +25°C, and +60°C). Measurements were carried during irradiation and for

P. A. Zimin, E. V. Mrozovskaya, V Yushkova, V. S. Anashin, P. A. Chubunov are with the Branch of JSC "United Rocket and Space Corporation" - "Institute of Space Device Engineering," 111250, Aviamotornaya st., 53, Moscow, Russia, e-mail: ZiminPA.msk@gmail.com

G. I. Zebrev, and P. A. Zimin, E. V. Mrozovskaya are also with the National Research Nuclear University MEPHI, Moscow, Russia.

20 hours thereafter at the same temperature, and after a year of storage at room temperature.

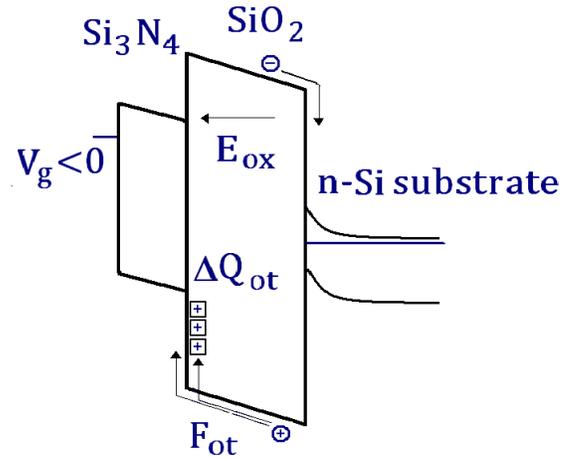

Fig. 1. The band diagram of the p-MNOS based RADFET.

The threshold voltage shift during irradiation was tracked at fixed output drain current. The reference drain current $I_D^{(r)}$, drain bias $V_D$ and the backward substrate bias $V_{sub}$ were chosen close to the "Zero Temperature Coefficient (ZTC) point" [5]: $I_D^{(r)} = 45$ μA, $V_D = -0.2$ V and $V_{sub} = +2.0$ V respectively. The operation gate voltage range was from 0 to - 12 Volts.

### B. Modeling of pMNOS based RADFETs

The charge buildup at the $Si_3N_4$-$SiO_2$ interface can be estimated as follows

$$\Delta Q_{ot} = q\, \eta_{eff}\, F_{ot} K_g t_{ox} D \,, \qquad (1)$$

where $D$ is a total dose, $q$ is the electron charge, $F_{ot}$ is the dimensionless hole trapping efficiency at the $Si_3N_4$–$SiO_2$ interface, $K_g \cong 8 \times 10^{12}$ cm$^{-3}$rad (SiO$_2$)$^{-1}$ is the electron-hole pair generation rate constant in the SiO$_2$, $t_{ox}$ is the silicon oxide thickness, $\eta_{eff}$ is the effective charge yield which depends on the oxide electric field $E_{ox}$, dose rate $P$, and irradiation temperature $T$ [6].

The threshold (reference) voltage shift due to the charge trapping at the $Si_3N_4$-$SiO_2$ interface is described as follows

$$\Delta V_T = -\Delta Q_{ot} t_N / \varepsilon_0 \varepsilon_N \ . \tag{3}$$

where $t_N$ is the Si$_3$N$_4$ layer thickness, $\varepsilon_0 \varepsilon_N$ is the nitride dielectric permittivity ($\varepsilon_N \cong 7.5$). Then, the sensitivity of a MNOS RADFET can be naturally defined as

$$A_D = \frac{|\Delta V_T|}{\Delta D} = q \frac{\eta_{eff}(E_{ox}) F_{ot} K_g t_{ox} t_N}{\varepsilon_0 \varepsilon_N} \tag{4}$$

The slopes of the dose curves are controlled by the charge trapping efficiency $F_{ot}$ and the effective charge yield $\eta_{eff}$. Equation (1) implies no annealing during irradiation. In fact, the dose dependence is slightly sub-linear due to temporal annealing (fading). The simplest approach to account for this effect is the rate equation

$$\frac{d\Delta V_T}{dt} = A_D P - \frac{\Delta V_T}{\tau_a}, \tag{5}$$

where $\tau_a = \tau_{a0} \exp(\varepsilon_a / kT)$ is the annealing time constant, $kT$ is the thermal energy, $\varepsilon_a$ is an activation energy. Eq. 5 leads to the exponential temporal dependencies which badly correspond to the long-term experimental behavior. It is well-known that the spread of the activation energies over a wide range could lead to a logarithmic temporal dependence.

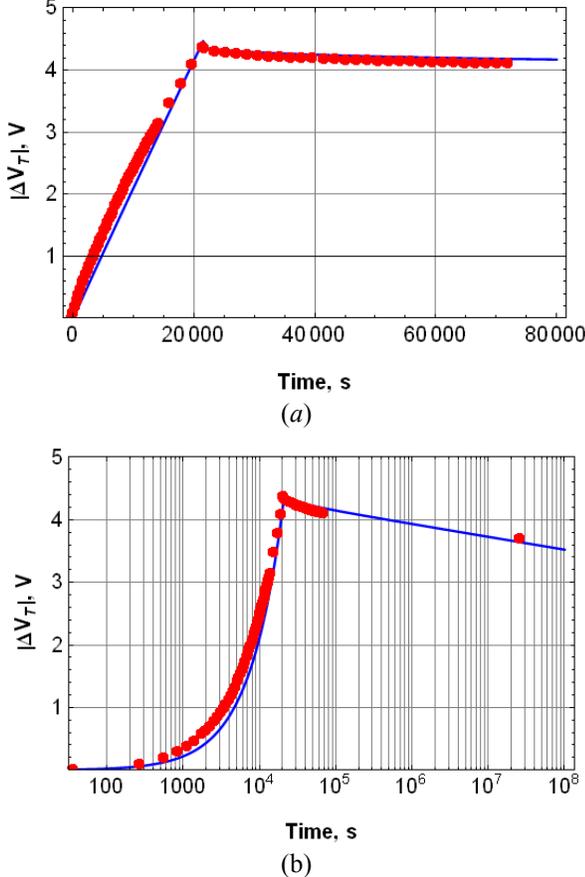

Fig. 2. Comparison of the experimental (points) and simulated (lines) threshold voltage shift during and after irradiation in linear (a) and logarithmic (b) temporal scales. Dose rate $P = 35$ rad(Si)/s, $t_{irr} = 2 \times 10^4$ s, irradiation temperature T = 60 °C, $b = 0.018$.

Particularly, the following relation for $\Delta V_T(D,t)$ was obtained in [7] considering the logarithmic annealing during and after irradiation

$$|\Delta V_T(t_1 \le t \le t_{irr})| = A_D\, D\left(1 - b\ln\left(\frac{t}{t_1}\right)\right) \qquad t_1 \le t \le t_{irr} \tag{6a}$$

$$|\Delta V_T(t > t_{irr})| = A_D D\left\{1 - b\left[\ln\left(\frac{t}{t_1}\right) + \frac{t - t_{irr}}{t_{irr}}\ln\left(\frac{t}{t - t_{irr}}\right)\right]\right\}, \qquad t > t_{irr} \tag{6b}$$

where $t_{irr}$ is the irradiation time, $t_1$ is the time of the first measurement, $b$ is a dimensionless fitting parameter, determined for the irradiation stage. Notice, that the model parameters $A_D$ and $b$ are not independent and determined by the choice of $t_1$. Particularly, the experimentally measured effective dose slope $A_D$ is a decreasing function of $t_1$, since a part of the trapped positive charge has at the time of the first measurement already annealed. As can be seen from (6), the constants at the different first measurement times $t_1$ and $t_2$ ($t_1 \le t_2$) are renormalized as follows [7],

$$A_D\left(1 - b\ln\left(\frac{t}{t_1}\right)\right) = \tilde{A}_D\left(1 - \tilde{b}\ln\left(\frac{t}{t_2}\right)\right), \tag{7}$$

where

$$\tilde{b} = \frac{b}{1 - b\ln(t_2/t_1)}, \quad \tilde{A}_D = A_D\left(1 - b\ln\left(\frac{t_2}{t_1}\right)\right). \tag{8}$$

Notice, that the parameter renormalization leaves invariant the product $A_D\, b = \tilde{A}_D \tilde{b}$, which turned out to be independent of the first measurement time.

The post irradiation relaxation (fading) can be described in another equivalent form

$$\frac{|\Delta V_T(t > t_{irr})|}{|\Delta V_T(t_{irr})|} = 1 - b^*\left[\ln\left(\frac{t}{t_{irr}}\right) + \frac{t - t_{irr}}{t_{irr}}\ln\left(\frac{t}{t - t_{irr}}\right)\right], \quad t > t_{irr} \tag{9}$$

where $b^*$ is the logarithmic slope that can be determined experimentally from the post-irradiation relaxation curve and linked with $b$ as follows

$$b^* = \frac{b}{1 - b\ln(t_{irr}/t_1)}. \tag{10}$$

The renormalization procedure excludes from consideration the unobservable values of maximum and minimum annealing times. At the same time, the renormalization introduces a dependence on the experimental conditions of measurements such as the time of first measurement. This is an inevitable consequence of the logarithmic form of temporal relaxation. The value of the physical constants depends on the scale that one chooses as the renormalization point.

### C. Experimental results

Experimental results for T= -40°C and T = +60°C are shown in Figures 2 and 3. These results are compared to the results of simulations carried out using the equations (6).

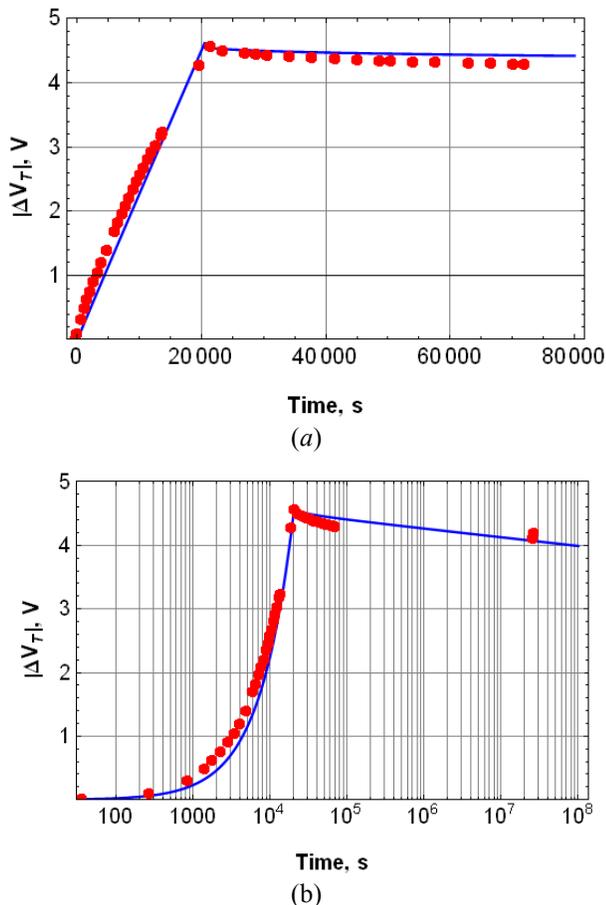

*(a)*

*(b)*

Fig. 3. Comparison of the experimental (points) and simulated (lines) threshold voltage shift during and after irradiation in linear (a) and logarithmic (b) temporal scales. Dose rate $P = 35$ rad(Si)/s, $t_{irr} = 2 \times 10^4$ s, irradiation temperature T = -40 °C, $b = 0.012$.

The post-irradiation annealing curves have apparently logarithmic temporal form. The rate of temporal dependence is slightly depends on temperature. This can be seen from a direct comparison of the annealing processes at different temperatures in Fig. 4.

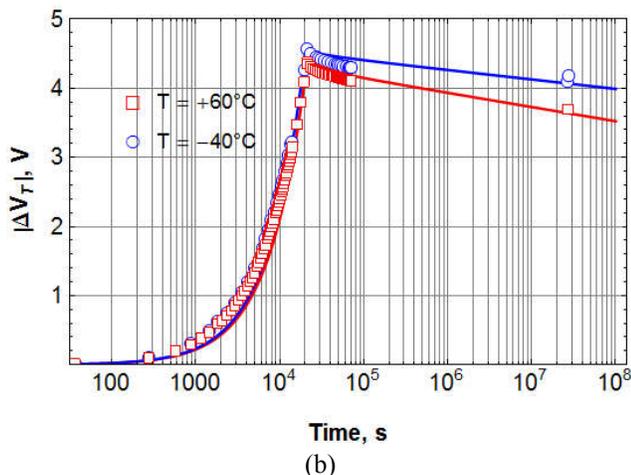

*(b)*

Fig. 4. Comparison of the experimental (points) and simulated (lines) results at different irradiation temperatures.

The results obtained demonstrate an occurrence of the logarithmic temporal dependencies in the pMNOS based RADFETs. We associate this effect with a large spread of the activation energies of the charged traps at the $Si_3N_4$–$SiO_2$ interface. The long-term annealing should be considered when calibrating and using the dosimeters.

## III. Conclusion

We have provided in this work a convincing experimental evidence of the logarithmic temporal annealing in the pMNOS based RADFETs. A quantitave model was proposed to describe this effect. It was found that the observed sublinear dose dependencies and the post-irradition annealing curves (fading) [8] can be simulated with the same parameters of the logarithmic annealing model. The renormalization procedure of the temporal parameters of the dose curves was validated.